# ARTICLE

# Interfacial Dzyaloshinskii-Moriya interaction in epitaxial W/Co/Pt multilayers

Sukanta Kumar Jena[a], Rajibul Islam[b], Ewelina Milińska[a*], Marcin M. Jakubowski[a], Roman Minikayev[a], Sabina Lewińska[a], Artem Lynnyk[a], Aleksiej Pietruczik[a], Paweł Aleszkiewicz[a], Carmine Autieri[b,c], Andrzej Wawro[a*]

Dzyaloshinskii-Moriya interaction (DMI) manifesting in asymmetric layered ferromagnetic films gives rise to non-colinear spin structures stabilizing magnetization configurations with nontrivial topology. In this work magnetization reversal, domain structure, and strength of DMI are related with the structure of W/Co/Pt multilayers grown by molecular beam epitaxy. Applied growth method enables fabrication of layered systems with higher crystalline quality than commonly applied sputtering techniques. As a result, a high value of $D$ coefficient was determined from the aligned magnetic domain stripe structure, substantially exceeding 2 mJ/m$^2$. The highest value of DMI value $D_{eff}$ = 2.64mj/m2 and  strength of surface DMI parameter $D_S$ = 1.83pJ/m for N=10 has been observed. Experimental results coincide precisely with those obtained from structure based micromagnetic modelling and density functional theory calculations performed for well-defined layered stacks. This high value of DMI strength originates from dominating contributions of the interfacial atomic Co layers and additive character from both interface types.

## Introduction

An asymmetric magnetic stack in which a thin magnetic layer (FM) is sandwiched between two heavy metals (HM) may host complex chiral magnetic structures. Interfacial Dzyaloshinskii-Moriya interaction (DMI), characterized by antisymmetric exchange energy term, is responsible for non-collinear alignment of interacting spins in a system missing inversion symmetry and exhibiting a strong spin-orbit coupling (SOC). Originally, it was proposed to explain observed weak ferromagnetism in antiferromagnets [1,2,3]. Later, in the B20 (space group P2$_1$3) bulk materials e.g. MnSi [4] and FeGe [5] with non-centrosymmetric structure, DMI was considered as an intrinsic property. In thin film layered structures DMI can be induced in a presence of broken inversion symmetry at HM/FM interface displaying strong SOC. As an effect, a non-collinear spin texture may emerge in a form of skyrmions, spin springs, or chiral domain walls (DW) [6,7]. Acquired equilibrium magnetic state of such layered structures results from a balance between anisotropy, exchange, DMI, and dipolar energies. Moreover, the width of emerging magnetic domains reflects an interplay between the competition of demagnetizing energy and domain wall energy in the multidomain state. Therefore, a proper setting of all energy components, including interfacial DMI enables tuning skyrmion sizes [7] or DW structure, which are a promising foundation for possible future solid state magnetic information technologies.

Methods for DMI energy controlling are mainly limited to choosing different materials for HM layers with opposite chirality of DMI of the bottom and top interfaces in HM$_1$/FM/HM$_2$ systems, or modification of their interfaces. In this scope different magnetic systems are investigated including even isolating component layers [8,9,10]. For Pt/Co/AlO$_x$ [11] the highest value of DMI $D_{eff}$= 1.71mJ/m$^2$ for Co(.6nm) and $D_S$=1.63 pJ/m. Where, $D_S$ is the surface DMI parameter $D_S$=$D_{eff}*t$, where the $t$ is the thickness of FM layer and $D_S$ is the independent on thickness of FM layer. Other isolating component i.e. MgO has been reported in literatures where the $D_S$= 1.58 pJ/m for Pt/Co/MgO systems. Boulle [8] et. al. reported that for Pt/Co/MgO, the value of $D_S$= 2.17 pJ/m which is higher than Pt/Co/AlO$_x$ and the total DMI strength arises from the both interface Pt/Co and Co/MgO which is same sign. However, fully metallic asymmetric layered stacks are investigated most frequently [12,13,14,15,16]. It is worth to mention that asymmetry in the crystalline structure of interfaces in chemically symmetric Pt/Co/Pt stacks [17] and Pd/Co/Pd multilayer [18] may also lead to substantial DMI. According to theoretical considerations [19,20] and review paper

a. Institute of Physics Polish Academy of Sciences, aleja Lotników 32/46, PL-02668 Warsaw, Poland
b. International Research Centre for Interfacing Magnetism and Superconductivity with Topological Matter, Institute of Physics Polish Academy of Sciences, aleja Lotników 32/46, PL-02668 Warsaw, Poland
c. Consiglio Nazionale delle Ricerche CNR-SPIN, UOS Salerno, I-84084 Fisciano (Salerno), Italy
*Corresponding authors:
Ewelina Milińska, Email: esieczko@ifpan.edu.pl
Andrzej Wawro, Email: wawro@ifpan.edu.pl

Electronic Supplementary Information (ESI): Crystallographic Information File, description of RHEED pattern, determination of DMI, details of micromagnetic simulation and details of DFT calculations.

[21], one of the highest strength of DMI is expected in Ir/Co/Pt systems. Both regular Ir/Co/Pt stack [7,22] and insertion of Ta layer to Co/Pt interface [23] were investigated reporting DMI strength as high as 2 mJ/m$^2$. One should mind that the Ir/Co/Pt system always does not give the additive DMI. Some experimental work [24] on Ir/Co/MgO and Pt/Co/MgO gives the same sign as a result of the DMI value is subtractive which contradicts the theoretically [19,20] study as well as experimental [7]. Ma [24] *et. al.* that at the interface Pt/Co and W/Co, the DMI sign is opposite which gives the additive DMI strength to W/Co/Pt system. However, it has been mention that the structure of W is either amorphous or polycrystalline.

Multilayer sputtered systems composed of Pt/Co/W stacks are another group studied intensively, recently. Such configuration is particularly energetically promising for skyrmion motion as the spin Hall angles at both interfaces exhibit opposite signs [25,26]. DMI influence and strength were determined by various methods. Jiang et al. [25] revealed the DMI strength to be equal to 1.5 mJ/m$^2$ which arises due to additive DMI at the both Pt/Co and Co/W interface and $D_S$ =1.5pJ/m by analytical considering a size dependence of the skyrmions stable at room temperature on magnetic field by means of Lorentz transmission electron microscopy (LTEM). Analysis of Pt/Co/M films (M – different metals) by Brillouin light scattering (BLS) showed that domain walls and spin waves are affected by the same strength of DMI, evidencing universality of this type of interaction in the case of different magnetic dynamics [27]. Determined DMI strength reported for Pt/Co/W stacks was found at the level of 1.3 mJ/m$^2$ ($D_S$=1.3pJ/m). DMI strength in the same stack determined through asymmetric domain wall velocity in in-plane magnetic field was reported to be much lower (0.19 mJ/m$^2$ and $D_S$=0.34pJ/m) [26]. Magnetic dead layer forming at the Co/W interface was shown to affect magnetic properties in Pt/Co/W multilayers with a lower repetition number [28]. Further growth of the structure improved interface quality and stabilized DMI strength at a constant value in the range between 0.6-1.2 mJ/m$^2$ depending on the Co layer thickness.

In this work, we investigate magnetisation reversal and magnetic domain structure in epitaxial W/Co/Pt multilayers with high perpendicular magnetic anisotropy (PMA). DMI strength is determined from the stripe domain structure aligned by in-plane applied field. We compare the investigated magnetic properties of the molecular beam epitaxy (MBE) grown layered structures with those usually fabricated by sputtering. The MBE technology results in a higher sample crystalline quality, crucial for high DMI strength. Moreover, we study an inverted sequence of component layers to those reported in other works [26,27,28]. In the case of MBE grown samples the sequence of constituent layers also substantially affects the crystalline structure of magnetic film and consequently the PMA. Simultaneously performed structure based micromagnetic simulations of the domain arrangement and density functional theory (DFT) calculations for various structural sample models were carried out to confirm the DMI strength determined from the experiment.

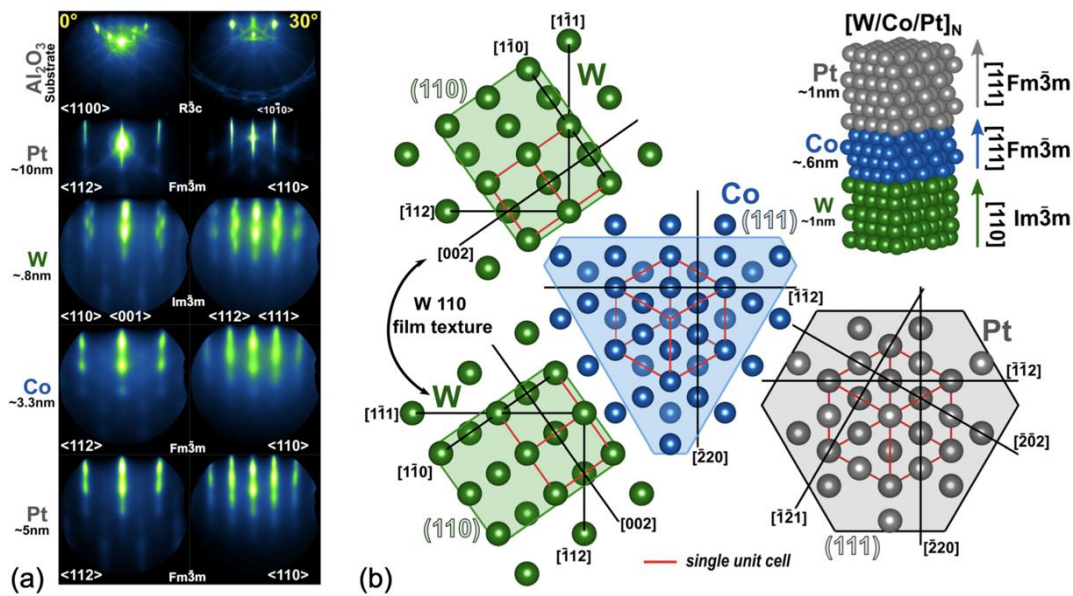

**Fig. 1 (a)** RHEED patterns corresponding to the sapphire substrate, Pt buffer, W, Co, and Pt cap layer. **(b)** Epitaxial relations between component films in [W/Co/Pt]$_N$ multilayer stack (shown in top-right corner) deduced from diffraction techniques. Atomic planes are marked with various background colouring for each component. Every scheme depicts perpendicular projection from the all atomic layers. Single unit cells in every layer are marked with the red lines. Mutual azimuthal orientations are shown by black lines, in the same reference axis system. The W crystallites are rotated randomly by 30° but only rotation by 90° results constructively in RHEED diffraction pattern shown in Figure 1 (a).

## Experimental

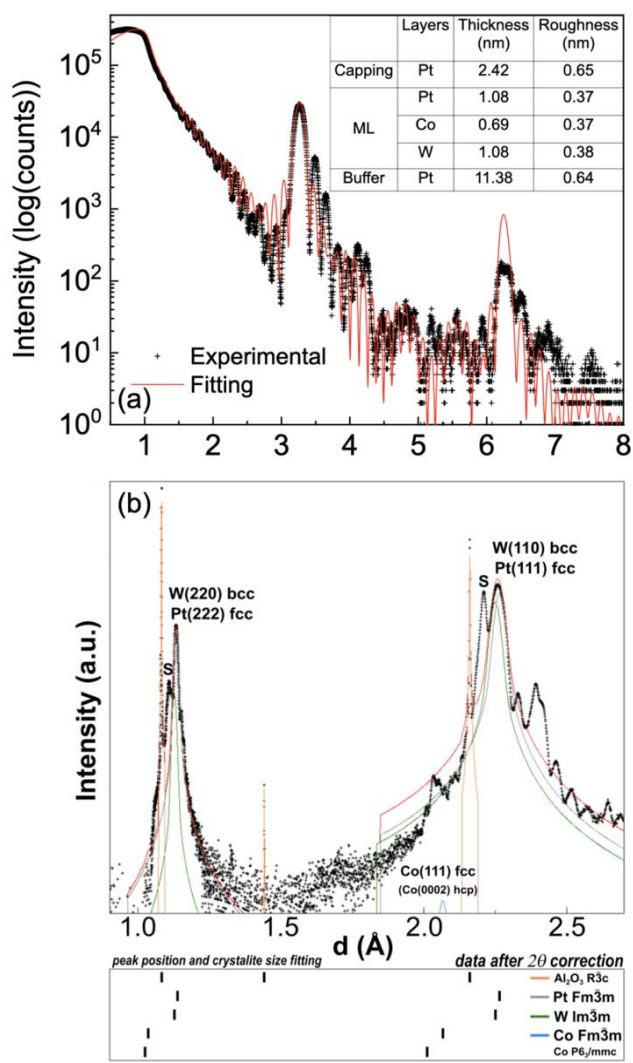

**Fig. 2** (a) Experimental XRR spectrum (black crosses) and the numerical fitting (red line) for the W/Co/Pt multilayer stack. Determined structural parameters of the multilayer (ML) are listed in the inserted table. (b) XRD pattern from W/Co/Pt multilayer. Its complex structure contains: main Pt and W peaks, superlattice peaks (S) and their higher orders, diffuse scattering, and Kato thickness fringes. Colour lines fit positions of the peaks (marked schematically in the bottom panel) from the individual component layers of the stack.

The investigated W/Co/Pt samples were grown in a molecular beam epitaxy (MBE) system operating in the base pressure lower than $8\times10^{-10}$ mbar in the form of [W(1)/Co(0.6)/Pt(1)]$_N$ multilayers (nominal thickness of individual layers is given in nm) with various number of the base W/Co/Pt stack repetition: **N** =10 and 20. The epitaxial growth was induced by (0001)-oriented sapphire substrate covered with 10 nm thick Pt(111) buffer. The stack was capped with an additional 2 nm thick Pt overlayer. Pt buffer was deposited at higher temperature of the substrate (750°C), whereas the subsequent multilayer at room temperature. Additionally, a reference W(10)/Co(3)/Pt(3) trilayer was grown in the same conditions as the multilayers. All materials were evaporated from the e-beam sources with rates at the level of 0.02 nm/s. The deposition rates and effective thicknesses were measured by a quartz-balance and monitored by XBS crossbeam mass spectrometer. The surface and epitaxial quality of the reference sample were observed by means of reflection high energy electron diffraction (RHEED). The structural ex-situ characterization of the samples has been performed by X-ray diffraction (XRD) (PANalitycal Empyrean X-ray diffractometer with Cu $K\alpha_1$, analysed using Maud [29] software) and X-ray reflectivity (XRR). Magnetization reversal of the samples in the magnetic field both applied in the plane of the sample and perpendicular direction was registered by superconducting quantum interference device (SQUID). Domain structure at the remanence state was investigated with use of magnetic force microscopy (MFM) operating in a lift mode. The all mentioned measurements were carried out at room temperature.

## Results

### A. Structure analysis

To monitor the crystalline structure of the W/Co/Pt layers RHEED patterns were recorded during the growth of the reference sample. The multilayered [W/Co/Pt]$_N$ samples were also studied ex-situ by XRD and XRR. The high quality of the layered structure has been confirmed by typical spectral features recorded by these methods.

Evolution of the RHEED pattern, recorded for a sequence of the component layers at various sample positions against the e-beam is shown in Fig. 1(a). Distinct streaks from every component layer taken along <110> and <112> azimuths (oriented in the sample plane) of the Pt buffer layer indicate that these layers grow as a good quality epitaxial textured film. Deduced the crystalline structure and interatomic spacing allowed resolving epitaxial relations at the interfaces (Fig. 1(b)). Visible atoms in the planar view belong to few atomic layers of each component film to illustrate more precisely the investigated structure. Horizontal, vertical, and oblique azimuth lines were plotted to clarify their mutual azimuthal orientations.

Fig. 2(a) shows a reflectometry curve for **N** = 20 multilayered sample, and iterative fitting allowing determination of the component layer thickness and the interface roughness. Two pronounced experimental Bragg peaks are clearly visible corresponding to the basic W/Co/Pt stack mean thickness. They are separated by numerous Kiessig fringes, which number corresponds to the basic trilayer repetition number building the whole stack. The shape of the fitting line is determined by the fitting parameters: component layer thickness and roughness of interfaces (listed in the inset table). These parameters were implemented in the optimal fitting of the experimental results and show that real values slightly exceed the nominal ones.

Fig. 2(b) depicts XRD pattern from multilayered W/Co/Pt structure, after data correction and rescaling to $d$ interplanar spacing. Apart from the strong signal from the sapphire substrate (sharp and highest peaks), the signal is dominated by the thick 10 nm Pt buffer layer. Experimental diffraction data was fitted using a constrained Rietveld refinement method (implementing known index peaks with fixed ratio for thin films, crystallite size, film thickness, and experimental setup parameters). Each fitted phase (cumulative signal from each element) is marked by colour line and indicator pointing centre of the phase peak. A peak corresponding to $[W/Co/Pt]_N$ superlattice resulting from the multilayered sample is located around $d$ = 0.22 nm ($2\theta$ = 40.8°) and was excluded from the fitting. The satellite peaks of higher orders and diffuse scattering are more pronounced in the lower $2\theta$ angle (higher $d$) shoulder (asymmetric satellites) of the tallest Pt (111) and the superlattice peaks, which indicates that the W/Co/Pt multilayer may have larger interplanar spacing at the bottom and top sides [30].

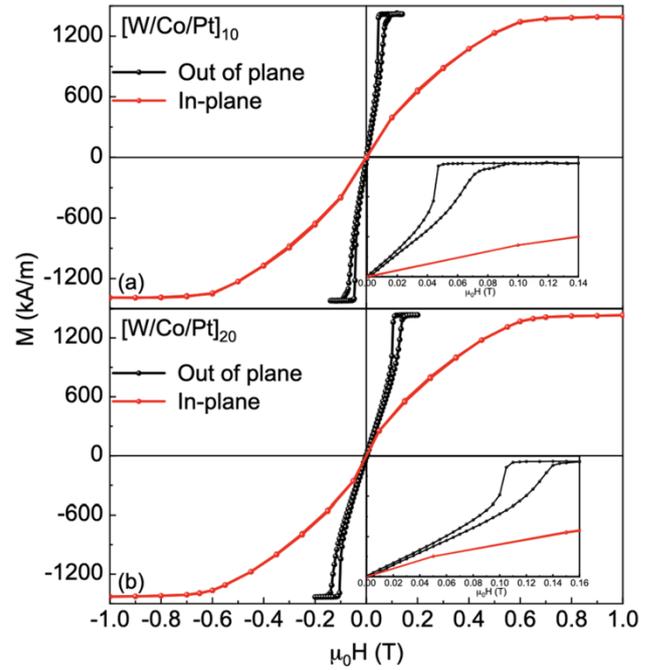

**Fig. 3** Magnetic hysteresis loops from: **(a)** $[W/Co/Pt]_{10}$ and **(b)** $[W/Co/Pt]_{20}$ multilayers measured in the field applied in the sample plane (red) and in perpendicular direction (black). Inserts show details of the loops in the low field range.

**Table 1** Structural parameters of W/Co/Pt stack: determined analytically by XRD, XRR and RHEED (rows 1-4) and implemented for basic stack models used in DFT calculation (rows 5-9).

|  | Crystalline structure | $a$ (Å) | $c$ (Å) | $\gamma$ (°) |
|---|---|---|---|---|
| Al2O3 substrate | $R\bar{3}c$ Trigonal | 4.754 | 12.982 |  |
| Pt | $Fm\bar{3}m$ fcc | 3.9237 |  |  |
| Co | $Fm\bar{3}m$ fcc | 3.5212 |  |  |
| W | $Im\bar{3}m$ bcc | 3.1875 |  |  |
| [W/Co/Pt] Model M1Co | $P1$ | 2.489 |  |  |
| [W/Co/Pt] Model M2wt | $P1$ | 2.694 |  | 120 |
| [W/Co/Pt] Model M3s | $P1$ | 2.707 | 33.149 |  |
| [W/Co/Pt] Model M4Pt | $P1$ | 2.774 |  |  |
| [W/Co/Pt] Model M5d | $P1$ | 2.707 |  | 120.197 |

Table 1 shows all phase mean parameters determined from RHEED analysis and XRR, XRD fittings. XRR and XRD provide out-of-plane symmetry information, whereas RHEED gives in-plane atom distances. Combination of these three methods allows collecting all necessary data for design a multilayer W/Co/Pt unit cell, scripted in Crystallographic Information File (CIF) (see: Supplementary Material). Different variations of this model implemented in density functional theory (DFT) calculations are discussed in the further part of this work.

**B. Magnetic properties**

Magnetization reversal in the multilayers $[W/Co/Pt]_N$ with repetition number **N** = 10 and 20 was investigated by SQUID.

Fig. 3 depicts hysteresis loops recorded in the magnetic field applied in the sample plane and in perpendicular direction. The loops are similar qualitatively for both trilayer repetition numbers **N**. The sample with **N** = 10 saturates easily in the perpendicular field of around 0.08 T (Fig. 3(a)). The estimated magnetization saturation is equal to 1420 kA/m, determined as for Co layers 0.69 nm thick, being confirmed by XRR measurements. With field decrease below 0.047 T, magnetisation signal drops suddenly reflecting nucleation of reversed domains. Due to field dependent magnetisation history the hysteresis opens at higher field range just prior to saturation. With further field, decrease reversed domains expand and at remanence a contribution of both types of local magnetization orientations becomes equal lowering the net magnetization signal down to zero (and coercivity as well). Such shape of the hysteresis suggests formation a domain structure magnetized perpendicularly to the film plane with magnetization vectors oriented in opposite directions.

Magnetization alignment lowers demagnetization energy of the whole system, even at the cost of energy of the emerging domain walls. This scenario is confirmed also by the closed shape of the loop measured in the field applied in the sample plane. It evidences that magnetization rotation is fully reversible and governed by a strong perpendicular anisotropy of the sample. At remanence no net in-plane magnetisation is observed confirming conclusions drawn from magnetization reversal in the perpendicular field. In-plane saturation is achieved in the field of 0.596 T. A difference in saturation fields in both measurement configurations allows determining the effective anisotropy coefficient discussed in the next section. As mentioned earlier M-H dependence recorded for the sample with **N** = 20 is qualitatively similar with saturation fields equal to 0.140 T and 0.607 T in the magnetic fields applied in perpendicular direction and in the sample plane, respectively (Fig. 3(b)).

**Table 2** Magnetic parameters describing [W/Co/Pt] multilayers. Symbol meaning is the following: $M_s$ - saturation magnetization, $H_k$ - anisotropy magnetic field, $W$ - domain width, $\Delta$ -domain wall width, $\sigma_{dw}$ - domain wall energy, $K_u$ - perpendicular anisotropy energy, $K_{eff}$ -effective anisotropy energy, $D$ - interfacial DMI strength, $D_S$ – surface DMI parameter and $D_{thr}$ - threshold interfacial DMI strength.

| N | $M_s$ (kA/m) | $H_k$ (T) | $W$ (nm) | $\Delta$ (nm) | $\sigma_{dw}$ (mJ/m$^2$) | $K_u$ (MJ/m$^3$) | $K_{eff}$ (MJ/m$^3$) | $D$ (mJ/m$^2$) | $D_S$ (pJ/m$^2$) | $D_{thr}$ (mJ/m$^2$) |
|---|---|---|---|---|---|---|---|---|---|---|
| 10 | 1420 | 0.595 | 108 | 4.37 | 3.56 | 1.68 | 0.442 | 2.65 | 1.83 | 0.39 |
| 20 | 1433 | 0.606 | 93 | 3.97 | 5.25 | 1.72 | 0.434 | 2.49 | 1.72 | 0.63 |

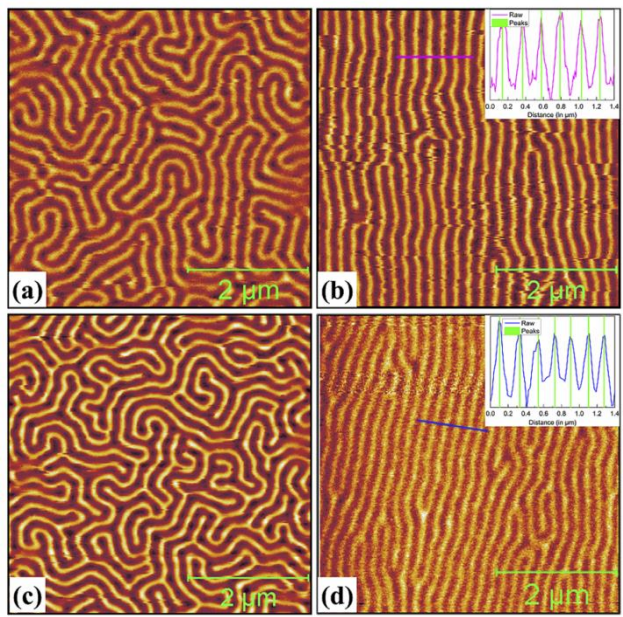

**Fig. 4** MFM images of the domain structures in the [W/Co/Pt]$_N$ multilayers with **(a, b)** N = 10 and **(c, d)** N = 20. **(a, c)** Labyrinth patterns correspond to demagnetized state. **(b, d)** Stripe domain alignment was obtained by demagnetization using in-plane oriented magnetic field. Inserts depict the profiles along the color section lines marked in the images with stripe domains.

Conclusions on magnetic domain structure drawn from the shapes of the hysteresis loops were fully confirmed by MFM measurements. Fig. 4 presents the MFM images of a domain structure at remanence in [W/Co/Pt]$_N$ multilayers with both **N** = 10 and **N** = 20. In as-deposited state, the samples exhibit a network of labyrinth domains magnetized in opposite directions perpendicularly to the sample plane (Fig. 4(a),(c)). Dark and bright colours in the image correspond to magnetization oriented up and down, respectively. The ratio of dark to bright areas in the images is equal to 1, which evidences the same number of domains oriented up and down and net magnetization vanishing at remanence.

After sample demagnetization by applying an in-plane ac field with decreasing amplitude after saturation, the labyrinth-like domains can be aligned into parallel stripe domain structure (Fig. 4(b),(d)). The width of the stripe domains was estimated from the profile lines shown as inserts. The domain width is determined as 108 nm and 93 nm in the samples with **N** = 10, and **N** = 20, respectively. Labyrinth-like domain width is slightly higher than that of stripe domains (Table 2). With increasing number of stacks in multilayers the domain width decreases.

## C. Dzyaloshinskii-Moriya interaction

DMI strength in layered structures may be analysed by various methods: e.g. by Brillouin light scattering (BLS) [23,11,31], spin wave propagation by VNA [32], and domain wall propagation induced by field [22,33] or electric current [27,34]. Description of DMI strength in multilayered systems from the domain structure size is favoured among numerous approaches commonly applied. One of the reasons is that spin wave propagation analysis in BLS spectra is significantly influenced by dipolar fields and the couplings between magnetic layers. Also, in the case of dense labyrinth-like magnetic domains being formed at low fields, the observation of the asymmetric domain wall propagation is difficult and may lead to erroneous estimation of DMI strength.

Determination of DMI strength in multilayered structures from domain size was proposed in several sequential approaches. A simplified model for a multilayered structure considered as an effective medium in which the whole film is treated as a single homogeneous magnetic layer was suggested by Woo et al. [35]. In this model, various magnetic parameters such as: magnetization, exchange constant, effective anisotropy, DMI strength have been properly scaled, because the effective medium has to reproduce the static and dynamic behaviour of a multilayer stack. An accurate analytical model of the stray field in materials with perpendicular anisotropy, taking into account the internal structure of the domain walls and their magnetostatic interactions between magnetic charges induced within these walls, that allows determining the DMI strength or exchange constant was proposed by Lemesh et al. [36]. In this work the mentioned earlier effective medium was tested successfully, providing accurate predictions. A similar approach of the effective medium was utilized for investigations of hybrid chiral domain walls in numerous multilayers with strong PMA and DMI [37]. Micromagnetic simulation performed with such approximation precisely reflected a twisted complex in-depth structure of the domain walls confirmed by circular dichroism in X-ray resonant magnetic scattering experiment.

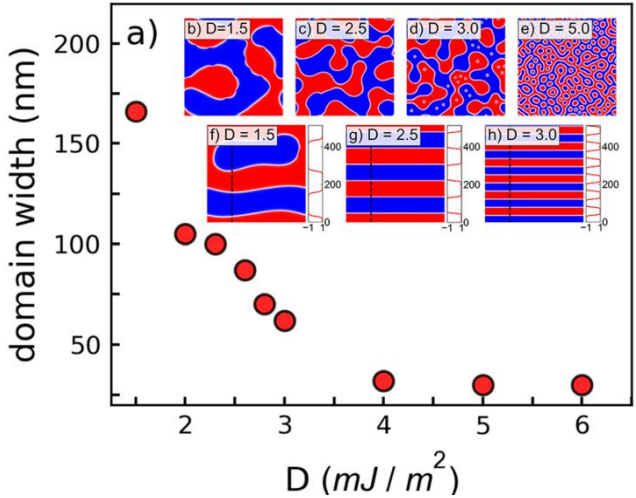

**Fig. 5** The domain pattern for various DMI strengths $D$ for [W/Co/Pt]$_{20}$ multilayer calculated from micromagnetic simulations. **(a)** The dependence of the domain width determined from the domain structures as a function of $D$ parameter. **(b-e)** Images (0.5 µm × 0.5 µm) of out-of-plane magnetized domains (red and blue colours stand for the up and down domains, respectively) for $D$ = 1.5, 2.5, 3.0, 5.0 mJ/m$^2$. **(f-h)** Relevant domain structures after applying the in-plane magnetic field, field, for $D$ = 1.5, 2.5, and 3.0 mJ/m$^2$. The map profiles (red) along the dashed lines (black) are presented on the right side.

Using the calculation scheme proposed by Legrand et al. [37], (details are presented in the Supplementary Material, Section 2) we assume that the configuration of parallel stripe domains observed by MFM at remanence corresponds to the minimum energy of investigated systems. All determined parameters implemented to the above-mentioned calculations are summarized in Table 2. The determined DMI strength is equal to $D$ = 2.65 mJ/m$^2$ and $D$ = 2.49 mJ/m$^2$, for $N$ = 10 and $N$ = 20, respectively, assuming $A$ = 13 pJ/m. For $A$ = 11 pJ/m the $D$ parameter takes a lower value of 2.41 mJ/m$^2$ and 2.18 mJ/m$^2$, for $N$ = 10 and $N$ = 20, respectively (not shown in Table 2). The value of $D_S$ = 1.72 pJ/m and 1.83 pJ/m for for $N$ = 10 and $N$ = 20, respectively which are the higher than the studied system with Pt/Co/W [25,26,38]. The sign convention of DMI strength $D$ has been taken as negative with respect to Pt/Co/W where the DMI strength $D$ is positive. The obtained values for multilayers with $N$ = 10 and $N$ = 20 are comparable showing that DMI strength in the investigated W/Co/Pt system does not depend substantially on the stack repetition number in multilayer samples, as was previously reported for Pt/Co/W system [28]. Performed calculations of the DMI coefficients threshold value $D_{trh}$ [36] (definition in the Supplementary Material, Section 2) return 0.39 mJ/m$^2$ and 0.63 mJ/m$^2$ for $N$ = 10 and $N$ = 20, respectively. Lower value of $D_{thr}$ compared to $D$ determined experimentally suggests that the structure of the domain walls might be of pure Néel type. However, from MFM images it is not possible to deduce a type of the domain walls.

**D. Micromagnetic simulations**

The domain patterns obtained from micromagnetic simulations (details given in Supplementary Materials, Section 3) exhibit domain size dependence on the DMI value. Fig. 5(b)-

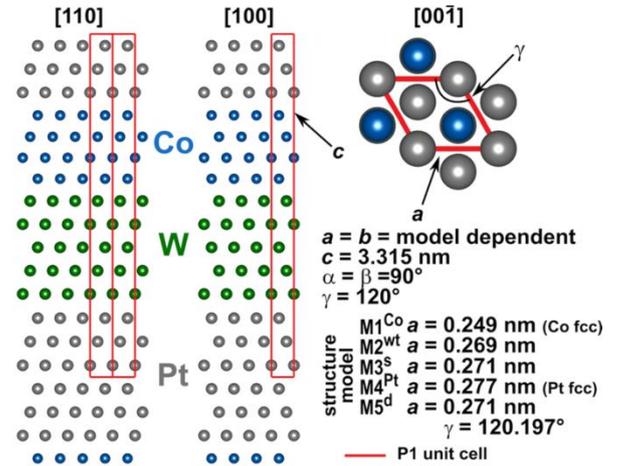

**Fig. 6** Different structure projections of the basic W/Co/Pt stack used for DFT calculations. P1 unit cell is marked with the red line.

(e) depict the static magnetic configurations of a multilayered thin film in $x$-$y$ plane ($z$ component of magnetization), where red and blue colours correspond to the oppositely magnetized domains with magnetization perpendicular to the film plane. The labyrinth-like pattern well resembles the MFM experimental results from as-deposited samples. The average domain sizes are estimated from fast Fourier transformation (FFT) analysis of the simulated images and compared with the cross-sections over a few domains in different areas of the image. The mean value of the width of the simulated domains decreases with increasing DMI strength (Fig. 5(a)). The best agreement between the simulated and the measured by MFM magnetic domain width is obtained for $D$ in the range between 2 mJ/m$^2$ – 2.5 mJ/m$^2$. Changes in this value range of $D$ parameter generate rather small influence on magnetic domain width, being in the range of 110 – 90 nm. The width analysis of the stripe domains forming after in-plane applying field with decreasing amplitude down to zero from $H$ = 0.65 T was also performed. The aligned domains for selected $D$ = 1.5, 2.5, and 3.0 mJ/m$^2$ are presented in Fig. 5(f)-(h). The decreasing size of domains with increasing $D$ is well visible. For $D$ = 2.5 mJ/m$^2$ the aligned domain size equal to 90 nm fits well with domain width determined from MFM images.

**E. Density functional theory calculations**

Experimentally determined DMI value was compared to the results of DFT calculations (details are provided in Supplementary Material, Section 4). As the repetition number of basic component stack interfaces does not influence substantially the DMI value (results shown in this work and reported in Ref. [28]), calculations were performed for a single basic component stack W/Co/Pt. Structural models were created on the basis of the growth analysis of individual layers in the reference sample. Lattice constants were taken from the analysis by RHEED and XRD (Fig. 1(a), 2(b) and Table 1).

Five structure models of the W/Co/Pt unit cell stack have been designed by preparing CIF (link to Supplementary Material) using VESTA [39] software. A primitive P1 space group unit cell was used to build the W/Co/Pt multilayer stack supercell. Fig. 6 shows three projections of this structure with marked P1 unit cell. The whole unit cell consists of 15 atomic layers: 5, 4 and 6 corresponding to W, Co and Pt, respectively (according to the nominal thickness). Taken unit cell parameter $c$ = 3.315 nm was extracted from fitting XRD superlattice (111) peak position ($d^S_{(111)}$ = 0.221 nm). Interatomic distances at the interfaces between different element atoms were taken as for equal-atomic alloy structures from the database [40] after volume isotropic correction. Each component layer has fcc-like crystal symmetry, but the whole unit cell is defined as a P1 space group. However, such assumption forces the change of W layer in-plane symmetry, because any unit cell cannot be described by more than one space group. In all considered below models the $c$ parameter is fixed, whereas variation of $a$ in-plane parameter reflects the main differences between them. In the first approach, the distances between in-plane atoms refer to experimentally determined extreme values (Co - M1$^{Co}$ model ($a^{Co}_{(110)}$ = 0.249 nm), Pt - M4$^{Pt}$ model ($a^{Pt}_{(110)}$ = 0.277 nm)). Three other intermediate models are considered to be closer to the real sample structure, where distances reflect experimentally acquired structure parameters of the multilayered sample. The M2$^{wt}$ model takes the phase-weighted mean spacing, $a^{wtMean}_{(110)}$ = 0.269 nm of each individual element (Pt, Co, W). In the M3$^S$ model $a^S_{(110)}$ = 0.271 nm is assumed, as determined from the superlattice XRD peak $d^S_{(111)}$, for the isotropic cell. Finally, in the M5$^d$ model, a small in-plane anisotropy strain was introduced to the M3$^S$ model, by change of the angle $\gamma$ = 120.197° in the P1 unit cell (Fig. 6). This deformation in [110] direction was observed by RHEED from the Co bare surface as varying streak spacing. Table 1 collects all sample parameters determined from XRD measurements and used for models described above.

DFT calculation results of DMI coefficient $D$ (details are provided in Supplementary Material, Section 4) for mentioned five different structural models of W/Co/Pt stack are plotted in Fig. 7(a). The values of $D$ are from the range of 2.04 - 2.57 mJ/m$^2$ with anticlockwise chirality [41]. It should be emphasized, however, that all the crystal structures are in satisfactory agreement with the experimental data. Furthermore, the atomic layer resolved DMI energy $d^k$ for the $k$-layer was calculated for the crystal structure M5$^d$ (Fig. 7(b)), as this structure is described by the highest degree of freedom, as can be expected in a real sample. The Co atoms belonging to the four atomic layers are labelled as Co$_1$, Co$_2$, Co$_3$ and Co$_4$. The 1$^{st}$ and 4$^{th}$ layers form the interfaces W/Co$_1$ and Co$_4$/Pt, respectively. Co$_2$ and Co$_3$ are the inner layers. The calculated value of $d^k$ shows that DMI is mostly located at two interfacial Co layers. The DMI at the W/Co and Pt/Co interface have opposite chirality. However, in the case of the opposite interfaces in the W/Co/Pt heterostructure, DMI vectors at these interfaces have the same chirality, resulting in the sum of two contributions and large DMI. The DMI strength from the inner layers is considerably smaller than from the interfacial layers. Chirality of 1$^{st}$, 2$^{nd}$ and 4$^{rd}$ layers are anticlockwise,

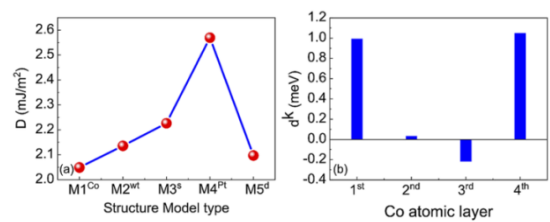

**Fig. 7 (a)** DMI strength calculated by DFT method for various structural models of W/Co/Pt stack (described in the text). **(b)** Distribution of DMI energy over individual atomic Co layers in M5$^d$ model of the W/Co/Pt stack.

whereas the chirality of 3$^{rd}$ layer is clockwise. Although Fig. 7(b) shows the distribution of the $d^k$ for crystal structure M5$^d$, other crystal structures exhibit the same qualitative behaviour. As a confirmation of the validity of our DFT results, the sum of the layer resolved DMI $\sum d^k$ equal to 1.85 meV is very close to the total DMI energy previously calculated for M5$^d$ model, which is equal to 1.80 meV according to the theoretical model.

## Discussion

Determination of DMI strength $D$ in epitaxial W/Co/Pt multilayered system was done using three independent approaches: an effective medium model based on the experimental results as well as numerical micromagnetic simulations and DFT calculations assuming perfect crystalline structure of the modelled system. Although these methods are entirely different, they reveal similar $D$ value substantially exceeding 2 mJ/m$^2$. Our studies confirm that this composition is energetically promising because both interfaces exhibit opposite signs.

In comparison to investigations of similar systems carried out by other groups, the obtained value is high. Reported $D$ coefficient values were from the range from 0.2 to 1.5 mJ/m$^2$ [26,27,28], however those systems were fabricated by sputtering technique and had reversed layer sequence, i.e. Pt/Co/W. It is commonly accepted and also shown in this work that DMI is interface sensitive – it depends on a chemical type and structural quality of interfaces with a magnetic component film. Crystalline quality of multilayers substantially depends on fabrication methods. Polycrystalline films with poorly defined structural relations at the interfaces are usually obtained by the sputtering method. On the contrary, the MBE method offers coherent growth of stack with high crystalline quality. In such systems, interfacial relations are much better defined. A crucial influence of deposition methods on DMI strength was presented in symmetrical Pt/Co/Pt system [22]. Moreover, epitaxial structures with mutually opposite component layer sequences are not equivalent to each other in the context of magnetic properties [42]. The bottom nonmagnetic layer affects considerably the growth type of the magnetic component. In the case of the W/Co/Pt system, a substantial crystalline symmetry and lattice parameter mismatches may occur at the W/Co interface. As a result, such configuration is expected to exhibit a lower PMA than the opposite sequence [25,26].

The high value of *D* reported in this work in comparison to other systems of the same type, however fabricated by sputtering, most probably arises from high quality of the investigated system and formed interfaces. It has been evidenced clearly by RHEED pattern, satellite peaks in XRD measurement, and Bragg peaks in the XRR curve. Experimental *D* value coincides well with those theoretically determined from numerical calculations performed for model structure with sharp and smooth interfaces. It has been reported that 25% interfacial mixing, which plays a role identical to interface roughness, reduces DMI strength by half [20]. In this context, slightly unexpected results were reported for [Co/Pd] MBE-grown multilayers [18]. High and increasing *D* value with **N** was correlated with interface roughness exceeding the thickness of the Co layer.

Presented here DFT calculations clearly show that DMI in W/Co/Pt system is a purely interfacial phenomenon. Dominating contribution is located at the interfacial Co layers. Obtained results are in good qualitative agreement with similar studies considering Pt/Co interface [20].

## Conclusions

Magnetization reversal, domain structure, and DMI were investigated in epitaxial W/Co/Pt multilayers. In comparison with similar sputtered systems, DMI strength estimated from magnetic domain structure is substantially higher and exceeds 2 mJ/m$^2$. This value is related to the opposite (additive) chirality of the DMI at the W/Co and Co/Pt interfaces and to the high quality of multilayer crystallinity confirmed by structural measurements. DMI strength determined in experimental way was confirmed by numerical micromagnetic modelling of the domain structure and DFT calculations performed for perfect structures with sharp and well-defined interfaces. Moreover, the DFT method reveals purely interfacial character of DMI with dominating contribution from Co atomic layers forming interfaces.

## Author Contributions

AW and EM proposed and supervised presented research. SKJ, AP, MJ and PA fabricated the samples and performed the RHEED measurements. SKJ and EM carried out MFM investigations. RM performed XRD and XRR measurement (incl. fitting). AL and SL were responsible for magnetic measurements by SQUID. MJ fitted and analyzed XRD, XRR and RHEED data and designed [W/Co/Pt] models. EM performed micromagnetic calculations using Mumax[3]. RI and CA conducted first principle calculation DFT. SKJ, MJ, EM and AW discussed and interpreted all results and prepared the manuscript.

## Conflicts of interest

Authors declare that they have no competing interests.

## Acknowledgements

This work was supported by the Foundation for Polish Science (FNP) under the European Regional Development Fund - Program [REINTEGRATION 2017 OPIE 14-20] and by Polish National Science Center projects: [2016/23/G/ST3/04196] and [2020/37/B/ST5/02299]. C. A. and R. I. were supported by the Foundation for Polish Science through the International Research Agendas program co-financed by the European Union within the Smart Growth Operational Programme. We acknowledge the access to the computing facilities of the Interdisciplinary Centre of Modelling at the University of Warsaw, [G73-23 and G75-10]. We acknowledge the CINECA award under the ISCRA initiative [IsC69 "MAINTOP"] and Grant, for the availability of high-performance computing resources and support [IsC76 "MEPBI"].

## References

1 T. Moriya, *Physical Review*, 1960, **120**, 91–98.
2 I. Dzyaloshinsky, *Journal of Physics and Chemistry of Solids*, 1958, **4**, 241–255.
3 T. Moriya, *Physical Review Letters*, 1960, **4**, 228–230.
4 S. Muhlbauer, B. Binz, F. Jonietz, C. Pfleiderer, A. Rosch, A. Neubauer, R. Georgii and P. Boni, *Science*, 2009, **323**, 915–919.
5 X. Z. Yu, N. Kanazawa, Y. Onose, K. Kimoto, W. Z. Zhang, S. Ishiwata, Y. Matsui and Y. Tokura, *Nature Materials*, 2011, **10**, 106–109.
6 W. Jiang, X. Zhang, G. Yu, W. Zhang, X. Wang, M. Benjamin Jungfleisch, J. E. Pearson, X. Cheng, O. Heinonen, K. L. Wang, Y. Zhou, A. Hoffmann and S. G. E. Te Velthuis, *Nature Physics*, 2017, **13**, 162–169.
7 C. Moreau-Luchaire, C. Moutafis, N. Reyren, J. Sampaio, C. A. F. Vaz, N. Van Horne, K. Bouzehouane, K. Garcia, C. Deranlot, P. Warnicke, P. Wohlhüter, J.-M. George, M. Weigand, J. Raabe, V. Cros and A. Fert, *Nature Nanotechnology*, 2016, **11**, 444–448.
8 O. Boulle, J. Vogel, H. Yang, S. Pizzini, D. De Souza Chaves, A. Locatelli, T. O. Menteş, A. Sala, L. D. Buda-Prejbeanu, O. Klein, M. Belmeguenai, Y. Roussigné, A. Stashkevich, S. Mourad Chérif, L. Aballe, M. Foerster, M. Chshiev, S. Auffret, I. M. Miron and G. Gaudin, *Nature Nanotechnology*, 2016, **11**, 449–454.
9 T. Srivastava, W. Lim, I. Joumard, S. Auffret, C. Baraduc and H. Beá, *Physical Review B*, 2019, **100**, 1–5.
10 A. Cao, X. Zhang, B. Koopmans, S. Peng, Y. Zhang, Z. Wang, S. Yan, H. Yang and W. Zhao, *Nanoscale*, 2018, **10**, 12062–12067.
11 M. Belmeguenai, J. P. Adam, Y. Roussigné, S. Eimer, T. Devolder, J. Von Kim, S. M. Cherif, A. Stashkevich and A. Thiaville, *Physical Review B - Condensed Matter and Materials Physics*, 2015, **91**, 1–4.
12 L. Camosi, S. Rohart, O. Fruchart, S. Pizzini, M. Belmeguenai, Y. Roussigné, A. Stashkevich, S. M. Cherif, L. Ranno, M. De Santis and J. Vogel, *Physical Review B*, 2017, **95**, 1–6.

13    D. A. Dugato, J. Brandão, R. L. Seeger, F. Béron, J. C. Cezar, L. S. Dorneles and T. J. A. Mori, *Applied Physics Letters*, , DOI:10.1063/1.5123469.

14    J. Lucassen, C. F. Schippers, M. A. Verheijen, P. Fritsch, E. J. Geluk, B. Barcones, R. A. Duine, S. Wurmehl, H. J. M. Swagten, B. Koopmans and R. Lavrijsen, *Physical Review B*, 2020, **101**, 1–6.

15    L. Wang, C. Liu, N. Mehmood, G. Han, Y. Wang, X. Xu, C. Feng, Z. Hou, Y. Peng, X. Gao and G. Yu, *ACS Applied Materials & Interfaces*, 2019, **11**, 12098–12104.

16    W. Legrand, D. Maccariello, N. Reyren, K. Garcia, C. Moutafis, C. Moreau-Luchaire, S. Collin, K. Bouzehouane, V. Cros and A. Fert, *Nano Letters*, 2017, **17**, 2703–2712.

17    A. W. J. Wells, P. M. Shepley, C. H. Marrows and T. A. Moore, *Physical Review B*, 2017, **95**, 1–8.

18    A. V. Davydenko, A. G. Kozlov, A. G. Kolesnikov, M. E. Stebliy, G. S. Suslin, Y. E. Vekovshinin, A. V. Sadovnikov and S. A. Nikitov, *Physical Review B*, 2019, **99**, 1–12.

19    A. Belabbes, G. Bihlmayer, F. Bechstedt, S. Blügel and A. Manchon, *Physical Review Letters*, 2016, **117**, 1–6.

20    H. Yang, A. Thiaville, S. Rohart, A. Fert and M. Chshiev, *Physical Review Letters*, 2015, **115**, 1–5.

21    A. Fert, N. Reyren and V. Cros, *Nature Reviews Materials*, , DOI:10.1038/natrevmats.2017.31.

22    A. Hrabec, N. A. Porter, A. Wells, M. J. Benitez, G. Burnell, S. McVitie, D. McGrouther, T. A. Moore and C. H. Marrows, *Physical Review B - Condensed Matter and Materials Physics*, 2014, **90**, 1–5.

23    K. Shahbazi, J. Von Kim, H. T. Nembach, J. M. Shaw, A. Bischof, M. D. Rossell, V. Jeudy, T. A. Moore and C. H. Marrows, *Physical Review B*, 2019, **99**, 1–13.

24    X. Ma, G. Yu, C. Tang, X. Li, C. He, J. Shi, K. L. Wang and X. Li, *Physical Review Letters*, 2018, **120**, 157204.

25    W. Jiang, S. Zhang, X. Wang, C. Phatak, Q. Wang, W. Zhang, M. B. Jungfleisch, J. E. Pearson, Y. Liu, J. Zang, X. Cheng, A. Petford-Long, A. Hoffmann and S. G. E. Te Velthuis, *Physical Review B*, 2019, **99**, 1–9.

26    T. Lin, H. Liu, S. Poellath, Y. Zhang, B. Ji, N. Lei, J. J. Yun, L. Xi, D. Z. Yang, T. Xing, Z. L. Wang, L. Sun, Y. Z. Wu, L. F. Yin, W. B. Wang, J. Shen, J. Zweck, C. H. Back, Y. G. Zhang and W. S. Zhao, *Physical Review B*, 2018, **98**, 174425.

27    D. Y. Kim, N. H. Kim, Y. K. Park, M. H. Park, J. S. Kim, Y. S. Nam, J. Jung, J. Cho, D. H. Kim, J. S. Kim, B. C. Min, S. B. Choe and C. Y. You, *Physical Review B*, 2019, **100**, 1–9.

28    I. Benguettat-El Mokhtari, A. Mourkas, P. Ntetsika, I. Panagiotopoulos, Y. Roussigné, S. M. Cherif, A. Stashkevich, F. Kail, L. Chahed and M. Belmeguenai, *Journal of Applied Physics*, , DOI:10.1063/1.5119193.

29    L. Lutterotti, *Nuclear Instruments and Methods in Physics Research Section B: Beam Interactions with Materials and Atoms*, 2010, **268**, 334–340.

30    J. Kanak, P. Wiśniowski, T. Stobiecki, A. Zaleski, W. Powroźnik, S. Cardoso and P. P. Freitas, *Journal of Applied Physics*, 2013, **113**, 023915.

31    H. T. Nembach, J. M. Shaw, M. Weiler, E. Jué and T. J. Silva, *Nature Physics*, 2015, **11**, 825–829.

32    J. M. Lee, C. Jang, B.-C. Min, S.-W. Lee, K.-J. Lee and J. Chang, *Nano Letters*, 2016, **16**, 62–67.

33    R. Soucaille, M. Belmeguenai, J. Torrejon, J. V. Kim, T. Devolder, Y. Roussigné, S. M. Chérif, A. A. Stashkevich, M. Hayashi and J. P. Adam, *Physical Review B*, 2016, **94**, 1–8.

34    L. Liu, X. Zhao, W. Liu, Y. Song, X. Zhao and Z. Zhang, *Nanoscale*, 2020, **12**, 12444–12453.

35    S. Woo, K. Litzius, B. Krüger, M. Y. Im, L. Caretta, K. Richter, M. Mann, A. Krone, R. M. Reeve, M. Weigand, P. Agrawal, I. Lemesh, M. A. Mawass, P. Fischer, M. Kläui and G. S. D. Beach, *Nature Materials*, 2016, **15**, 501–506.

36    I. Lemesh, F. Büttner and G. S. D. Beach, *Physical Review B*, 2017, **95**, 1–16.

37    W. Legrand, J. Y. Chauleau, D. Maccariello, N. Reyren, S. Collin, K. Bouzehouane, N. Jaouen, V. Cros and A. Fert, *Science Advances*, , DOI:10.1126/sciadv.aat0415.

38    I. Benguettat-El Mokhtari, A. Mourkas, P. Ntetsika, I. Panagiotopoulos, Y. Roussigné, S. M. Cherif, A. Stashkevich, F. Kail, L. Chahed and M. Belmeguenai, *Journal of Applied Physics*, 2019, **126**, 133902.

39    K. Momma and F. Izumi, *Journal of Applied Crystallography*, 2011, **44**, 1272–1276.

40    A. Jain, S. P. Ong, G. Hautier, W. Chen, W. D. Richards, S. Dacek, S. Cholia, D. Gunter, D. Skinner, G. Ceder and K. A. Persson, *APL Materials*, 2013, **1**, 011002.

41    S. Emori, U. Bauer, S. M. Ahn, E. Martinez and G. S. D. Beach, *Nature Materials*, 2013, **12**, 611–616.

42    A. Wawro, Z. Kurant, M. Tekielak, P. Nawrocki, E. Milińska, A. Pietruczik, M. Wójcik, P. Mazalski, J. Kanak, K. Ollefs, F. Wilhelm, A. Rogalev and A. Maziewski, *Journal of Physics D: Applied Physics*, , DOI:10.1088/1361-6463/aa6a94.